\documentclass[10pt,twocolumn,journal]{IEEEtran}

\usepackage[T1]{fontenc}

\usepackage{amsmath}
\usepackage{setspace}
\usepackage{amssymb}
\usepackage{stfloats}
\usepackage{cite}
\usepackage{ragged2e}
\usepackage[ruled,vlined,linesnumbered]{algorithm2e}
\usepackage{amsfonts}
\usepackage{mathrsfs}
\usepackage{amsmath,amsthm}
\usepackage{array,booktabs}
\usepackage{subfigure}
\usepackage{multirow}
\usepackage{cuted}
\usepackage{multicol}
\usepackage{graphicx}
\usepackage{subfigure}
\usepackage{graphicx,xcolor,bm}
\usepackage{hyperref}
\usepackage{threeparttable}
\usepackage{dcolumn}
\usepackage{setspace}
\usepackage{makecell}
\usepackage{lipsum}
\usepackage{enumerate}
\pagecolor[rgb]{0.9, 0.99, 0.9}

\usepackage{cite,bm}
\graphicspath{{figures/}}
\begin{document}

\title{Accelerating the Delivery of Data Services over Uncertain Mobile Crowdsensing Networks}

\author{Minghui Liwang, \IEEEmembership{Member, IEEE}, Zhipeng Cheng, \IEEEmembership{Member, IEEE}, Wei Gong, Li Li Yuhan Su, Zhenzhen Jiao, Seyyedali Hosseinalipour, \IEEEmembership{Member, IEEE}, Xianbin Wang, \IEEEmembership{Fellow, IEEE}, Huaiyu Dai, \IEEEmembership{Fellow, IEEE}
\thanks{Minghui Liwang and Yuhan Su are with School of Informatics, Xiamen University, China. Zhipeng Cheng is with Soochow University, China. Wei Gong and Li Li are with Tongji University, China. Zhenzhen Jiao is with the iF-Labs, Beijing Teleinfo Technology Co., Ltd., CAICT, China. Seyyedali Hosseinalipour is with University at Buffalo, SUNY, USA. Xianbin Wang is with Western University, Canada. Huaiyu Dai is with NC State University, USA.
}
}

\maketitle
\newcommand\blfootnote[1]{%
\begingroup
\renewcommand\thefootnote{}\footnote{#1}%
\addtocounter{footnote}{-1}%
\endgroup
}
\begin{abstract}
The challenge of exchanging and processing of big data over \underline{m}obile \underline{c}rowd\underline{s}ensing (MCS) networks calls for designing  seamless data service provisioning mechanisms to enable utilization of resources of mobile devices/users for crowdsensing tasks. Although conventional onsite spot trading of resources based on real-time network conditions can facilitate data sharing, it often suffers from prohibitively long service provisioning delays and unavoidable trading failures due to requiring timely analysis of dynamic network environment. These limitations motivate us to investigate an \underline{i}ntegrated \underline{f}orw\underline{a}rd and \underline{s}pot \underline{t}rading mechanism (iFAST), which entails a novel hybrid data trading protocol with time efficiency, over uncertain MCS ecosystems. In iFAST, the \textit{sellers} (i.e., mobile devices who can contribute data) can provide \textit{long-term} or \textit{temporary} sensing services to the \textit{buyers} (i.e., sensing tasks). Specifically, it enables signing long-term contracts in advance of future transactions through a forward trading mode, via analyzing historical statistics of the network/market, for which the notion of \textit{overbooking} is introduced and promoted. iFAST further encourages the buyers with unsatisfying service quality to recruit temporary sellers through a spot trading mode, considering the current network/market conditions. We analyze the fundamental blocks of iFAST and provide a case study to demonstrate its performance. Inspirations for future research directions of next-generation sensing and communication are summarized. 
\end{abstract}

\begin{IEEEkeywords}
Mobile crowdsensing, forward and spot trading, hybrid service provisioning, overbooking, uncertain networks
\end{IEEEkeywords}

\section{Preliminary and Motivation }

\IEEEPARstart{T}{he} past decade has witnessed an explosive growth in smart mobile devices embedded with growing sensing and computing capabilities, leading to a wide range of big data-driven applications (e.g., transportation information gathering and environment monitoring)~\cite{1,2}. Accordingly, new architectures have been developed to provide a platform for the implementation of such applications, among which, \underline{m}obile \underline{c}rowd\underline{s}ensing (MCS) networks are of paramount importance. An MCS network orchestrates the powerful computing/communication/storage capabilities of geo-distributed mobile users with available resources and capabilities, to conduct sensing and data collection/sharing within multiple interested regions (i.e., point of interest, PoI)\cite{3}. 


One primary obstacle in the implementation of MCS networks is the generation/collection, storage, and transmission of large volume of data, which can impose additional \textit{cost} (e.g., data transfer costs and thus mobile phone bills) on selfish mobile devices~\cite{4,5}. 
A practical approach is to provide monetary incentives to engage them~\cite{2,6,7}, which draws a resemblance between MCS networks and \textit{data trading markets} by involving a series of \textit{transactions}. Specifically, in each transaction, \textit{service buyers} (e.g., task owners) will make payments to \textit{service sellers} (i.e., mobile users) for their contributions, with an MCS coordinator (i.e., an edge server) supervising the trading procedure. A fundamental problem in such markets is to \textit{find proper matching between buyers and sellers, while considering task- and device-specific factors, e.g., quality of service and service price}. 

Existing works have mainly focused on data trading (i.e., buyer-seller assignment and incentive designs) over MCS networks through either online or offline mode. The former is a form of \textit{spot trading}, where decisions are made by analyzing the current network/market situation, e.g., current wireless channel qualities and data supply/demand. Nevertheless, its implementation suffers from the following drawbacks~\cite{8,9}:

\begin{figure*}[h!t]
\centerline{\includegraphics[width=0.999\linewidth]{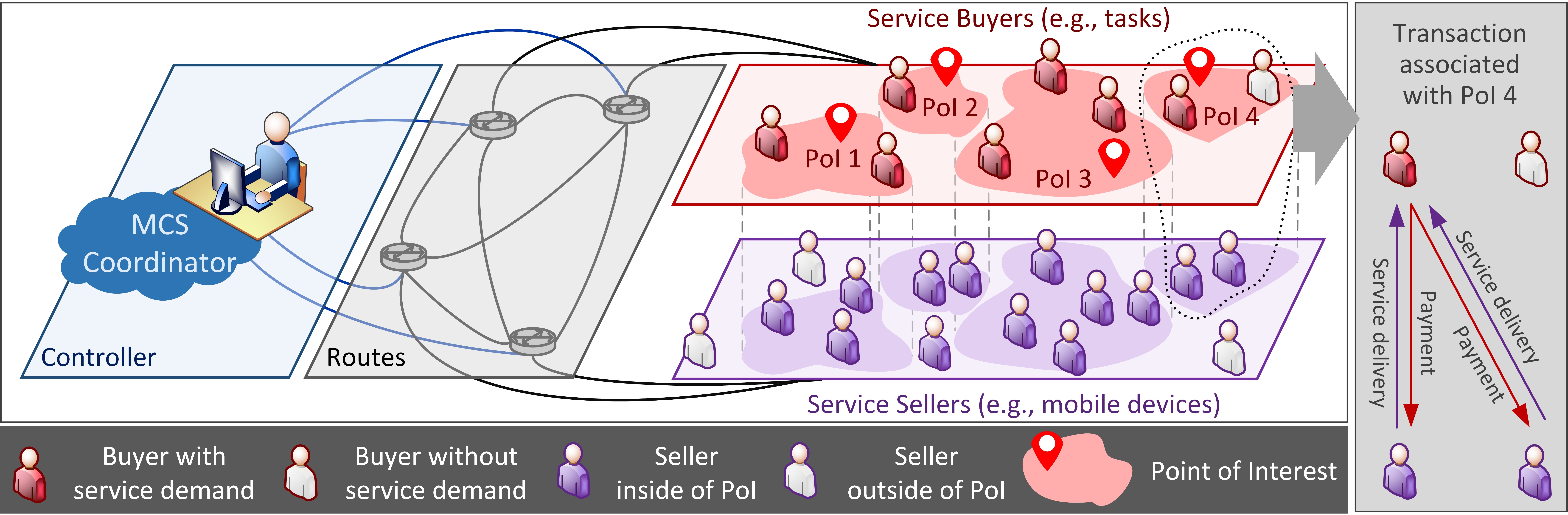}}
\caption{A schematic of service trading over MCS networks, which consists of an MCS coordinator that supervises the market, multiple buyers and sellers. In each transaction, each seller within the corresponding PoI of a buyer can provide sensing services to that buyer, while receiving a certain payment.}
\label{fig1}
\end{figure*}

\noindent
$\bullet$~\textit{Low time efficiency}. Spatio-temporal variations of MCS networks, e.g., varying users' locations and wireless channel qualities caused by the mobility of devices, incurs extra latency in decision-making during each transaction. For example, buyers and sellers have to negotiate a trading agreement, while simultaneously analyzing the current network/market situation. This online decision-making process can \textit{i)} degrade the value (e.g., freshness) of sensed/collected data of deployed devices, and \textit{ii)} reduce the amount of time allocated to practical data delivery, and thus imposing a low time efficiency. Moreover, a spot trading decision is only suitable for the current market/network situation, leading to accumulation of latency upon handling a large number of online transactions over time. 

\noindent
$\bullet$~\textit{Excessive energy consumption}. The repeated decision-making procedure further imposes high energy consumption, especially for battery- and power-limited mobile devices. Such an online decision-making can subsequently lead to excessive carbon emissions, hindering the sustainable development of MCS networks.  

\noindent
$\bullet$~\textit{Service failures}. During a spot transaction, each buyer is risking the failure to receive its desired data service after spending a certain period of time on negotiating an agreement, due to the limited and dynamic resource supply~\cite{8,10}.

The above-mentioned drawbacks of spot trading have attracted research on offline data trading over MCS networks, which refer to as \textit{forward service trading mode}. Offline decision-making can be conducted ahead of future transactions through analyzing the historical statistics of the market/network, which leads to a higher time/energy efficiency. Moreover, the impact of service failures can be mitigated, as participants are not required to make trading decisions onsite \cite{8}. Nevertheless, forward trading may suffer from an unsatisfying trading efficiency when the available statistical information cannot well capture the instantaneous situation of the market/network (e.g., due to users' mobility and abrupt environmental changes).
To address this limitation, most existing works that focus on offline trading design for MCS networks consider some form of prior knowledge of the network, e.g., the known knowledge of data quality, which, however, is impractical in many real-world environments~\cite{11}. 

Considering the above-discussed advantages and disadvantages of different trading modes, we are motivated to develop responsive, reliable, and yet practical data service trading mechanisms for MCS ecosystems under uncertainties (e.g., dynamic resource supply/demand, unpredictable mobility patterns of devices, time-varying wireless channels). To this end, we propose a novel trading paradigm called \underline{i}ntegrated \underline{f}orw\underline{a}rd and \underline{s}pot \underline{t}rading (iFAST), which accelerates data provisioning especially in dynamic MCS environments, through exploiting both offline and online trading mechanisms. It further introduces \textit{overbooking}~\cite{12} and promotes its utilization. In particular, the notion of overbooking exists in other domains such as in airline systems, where some airlines routinely overbook flight tickets since the passengers may miss their flights or change their schedules; otherwise, each flight may take off with 15\% vacant seats on average, causing economic losses. In iFAST, overbooking provides a buyer with the option of overbuying resources exceeding its actual demand, while enabling each seller to overbook resources surpassing its local supply, to cope with the dynamics of service demand and supply. For instance, a buyer can prepay for 5 sellers although it only requires 3 of them, in case some prepaid sellers leave the market. In this article, the word “trading” is a general economic term for data provisioning; while “transaction” is used to indicate a specific trading event, in which agreements between buyers and sellers are made.



\begin{figure*}[t!]
\centerline{\includegraphics[width=0.999\linewidth]{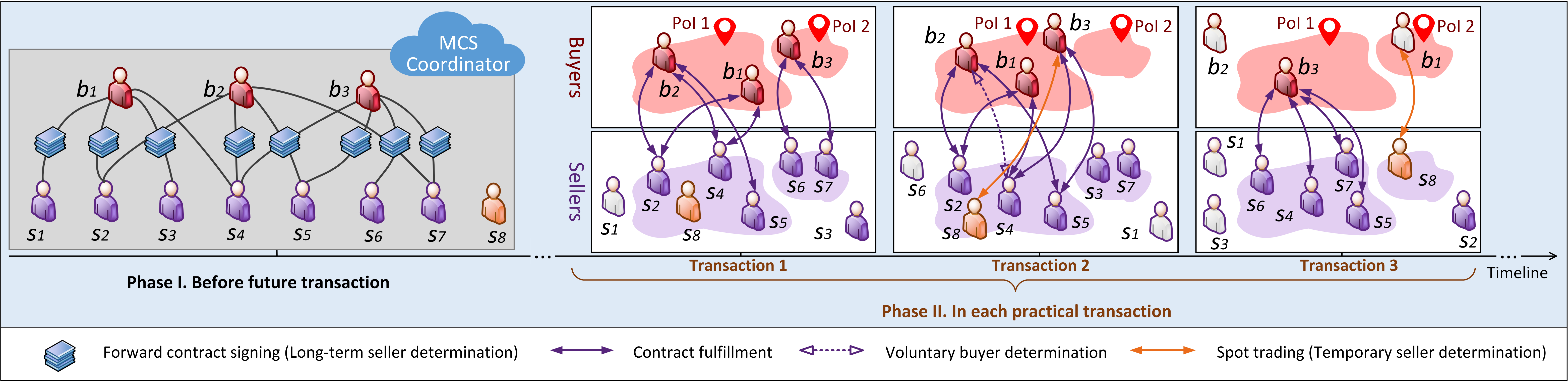}}
\caption{Timeline and transaction examples of iFAST. Forward contracts are signed among a portion of the buyers and sellers (sellers $s_1-s_7$ are long-term sellers) in advance of future transactions, and will be fulfilled during each practical transaction. During each transaction, $s_8$ can be recruited as a temporary seller, when any buyer’s desired service quality has not been reached, since some long-term sellers may have left the PoI.}
\label{fig2}
\end{figure*}

\section{iFAST for Uncertain MCS Networks}

\subsection{Overview}
A schematic of our interested market is depicted in Fig. 1, comprising three main components: \textit{i)} an MCS coordinator (e.g., edge server) with multiple point of interest (PoI), where each PoI covers a region that service buyers are interested in its data, \textit{ii)} buyers (e.g., sensing task owners), and \textit{iii)} sellers (e.g., smart mobile devices equipped with sensing units). Buyers and sellers are geo-distributed and can move among different PoIs. 

The MCS coordinator acts as a central controller responsible for handling data requests of buyers and assigning their sensing tasks to proper sellers. Besides, buyers periodically publish their requests for acquiring sellers interested in the execution of their sensing tasks with the help of the MCS coordinator, and sellers respond to those requests while receiving certain monetary incentives. More importantly, we then investigate the existing uncertainties that reveal the random and unpredictable nature of the considered data service trading market: 

\noindent
$\bullet$~\textit{Dynamic service demand/supply}. Service (i.e., data) demands and supplies are dynamic due to the mobility and personal preference (e.g., selfishness) of mobile devices. For instance, some buyers may not always take part in a transaction, e.g., a buyer who has already obtained the required data from other sources prior to a transaction will not buy the services it was supposed to. Moreover, a seller who moves outside of the target PoI can not contribute data to the sensing tasks within that PoI. 

\noindent
$\bullet$~\textit{Time-varying communication environment}. Since data services are requested and delivered through wireless communication networks, the temporal variation of wireless channel conditions can have a considerable impact on the efficiency of tasks.  

\noindent
$\bullet$~\textit{Unwarranted service quality}. The quality of the sensing data (and thus services) provided by each seller is constrained by multiple factors. For instance, its limited resources (e.g., communication, computation, storage) and capability (e.g., hardware settings of embedded sensors and the processing unit). Also, sellers may have their own resource-consuming local tasks. For example, during a transaction, a seller may postpone the processing of the assigned sensing tasks after its local tasks have been completed. Consequently, fluctuations and unpredictability in service quality are anticipated within MCS networks. Furthermore, the variability in channel conditions over time can also introduce uncertainties such as significant data transmission delay.

\begin{figure}[b!]
\centering
\centerline{\includegraphics[width=0.999\linewidth]{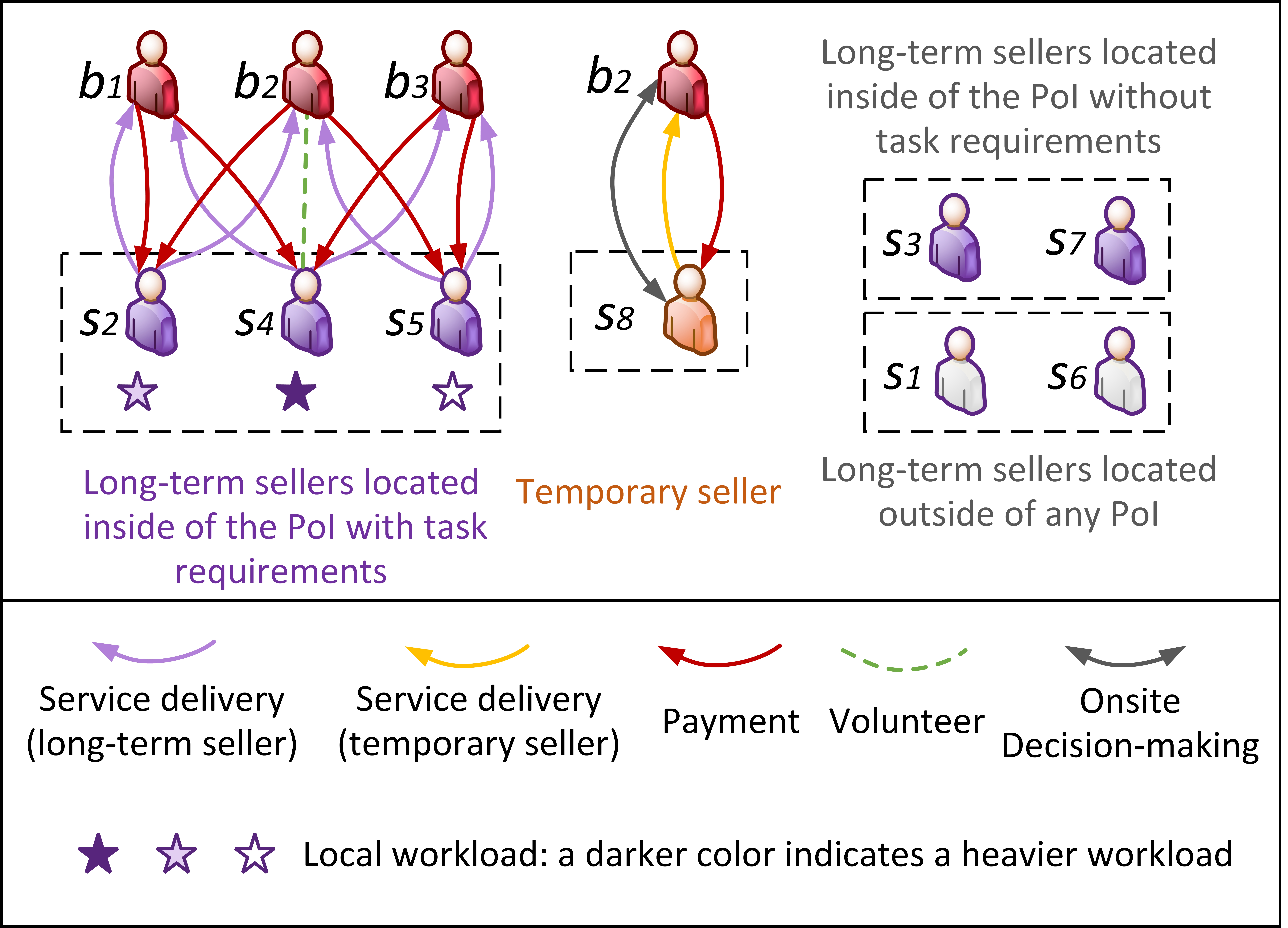}}
\caption{Interactions among sellers and buyers associated with Transaction 2 in Fig. 2. Long-term sellers $s_2, s_4, s_5$ can offer services to the buyers according to the pre-signed contracts, while buyer $b_2$ is selected as a volunteer of $s_4$. Besides, since the desired service quality of $b_2$ may not be satisfied, $s_8$ is determined as a temporary seller that provides one-time data service in the current transaction. Long-term sellers $s_1, s_3, s_6, s_7$ are unavailable during this transaction due to their locations.}
\end{figure}

To accelerate the cost-effective data provisioning while taking into account for the uncertainties in MCS networks, our methodology, iFAST, integrates forward and spot trading, introducing a phase-wise trading mechanism (see Fig. 2), where the the former phase of iFAST (i.e., forward trading) captures the time period before practical transactions, while its second phase (i.e., spot trading) encompasses the time duration of each transaction. 
In particular, forward trading phase enables each buyer to pre-negotiate trading agreements with a certain set of sellers (e.g., with terms such as price and the expected data quality), via analyzing historical statistics of the network/market. The employed sellers will subsequently serve their respective buyers for multiple transactions. We refer to the employed sellers during the first phase as \textit{long-term sellers}, while the corresponding trading agreements are referred to as \textit{forward contracts}. Moreover, spot trading phase is seen as a backup plan for the buyers who are suffering from unsatisfying service quality during each transaction (e.g., absence of long-term sellers). Specifically, buyers can employ a certain number of sellers without forward contracts to meet their service quality requirements based on the current network/market condition. We refer to the sellers employed in this phase as \textit{temporary sellers}. Different from long-term sellers, temporary sellers only offer services to their respective buyers during the current transaction (i.e., one-time services). Integration of both long-term sellers and temporary sellers in iFAST framework ensures responsive and seamless data provisioning to buyers.

\subsection{Key Challenges in iFAST}
The efficient implementation of iFAST over real-world MCS networks can confront a variety of challenges. Specifically, deploying the forward trading faces the following key challenges. 

\noindent
$\bullet$~\textit{Reaching mutually beneficial forward contract terms}. By entering into forward contracts with selected sellers, each buyer gains the following rights during each transaction: \textit{i)} paying a pre-determined/fixed price to each long-term seller rather than a fluctuant one, and \textit{ii)} executing its sensing service with a certain quality assurance offered by each long-term seller that \textit{attends} a transaction. Also, each long-term seller can receive reasonable pre-negotiated/guaranteed payments during each transaction. Besides, obligations should be considered as a penalty when a seller fail to offer the pre-determined services; or when a buyer is unwilling to buy the contractual services. Designing each forward contract thus entails reaching a mutually beneficial consensus between sellers and buyers, which is non-trivial since it generally represents a multi-objective optimization problem with complicated constraints. To address this challenge, game theoretic equilibriums, auction-based agreements, and bilateral/multilateral negotiations are feasible methodologies. 

\noindent
$\bullet$~\textit{Designing optimum overbooking rate}. The dynamics of service demand/supply represents one of the important features of data trading market, for which we introduce the notion of \textit{overbooking} that can improve the robustness of service provisioning against uncertainties in resource demand/supply. In iFAST, overbooking allows each seller’s promissory resources (services) to be more than its actual resource supply (the overbooking rate of each seller is calculated by the percentage of booked resources that exceeds its resource supply) to avoid having underutilized resources. For example, a seller can agree to offer more resources than its actual local resources in case some buyers are absent in future transactions. 
Moreover, overbooking allows each buyer to acquire more services (resources) than its actual demand, to take into account for dynamics of service supply (the overbooking rate of each buyer is defined by the percentage of booked services that exceeds its demand). For example, although the services provided by three long-term sellers are sufficient to meet the demand of a buyer, it may recruit more sellers to cope with the absence of some sellers with contracts during practical transactions (see examples in Fig. 2). Thus, involving overbooking in data trading can potentially improve resource utilization rate and bring long-term profits to both parties. However, since forward contracts are obtained before future transactions and can only be determined by analyzing historical statistics, the design of a good overbooking rate is challenging. For instance, a large overbooking rate on the seller's side can result in service failures , e.g., some contractual buyers may be unable to receive the required resources since the long-term seller fails to provide the overall resource demands by its limited resources. Also, a small overbooking rate may bring excessive latency to onsite decision-making, possible trading failures, and resource underutilization. 

\noindent
$\bullet$~\textit{Managing potential risks}. Although forward trading mode may lead to opportunistic profits for sellers and buyers, possible losses can also result from the improper design of contract terms, resulting in the coexistence of risks and opportunities. In iFAST, risks involve three key forms. 

\begin{enumerate}
\item Unsatisfying participant’s utility: Each seller/buyer faces the risk of obtaining an undesired utility due to the uncertainties of the market. For example, a buyer may receive a low service quality during a transaction since some long-term sellers are unable to provide services. Besides, a long-term seller is risking possible economic losses, especially with heavy local workload.

\item Undesired service failure: Participants in dynamic MCS networks are facing the risk of service failures. For instance, a high overbooking rate could lead to situations where buyers are unable to receive the services they were guaranteed, owing to potential resource constraints of long-term sellers.

\item Unsatisfying resource utilization: For long-term sellers, maintaining a low overbooking rate could result in the underutilization of resources, leading to economic losses, especially in scenarios where a significant number of buyers do not participate in a transaction. 

\item Over-budget risk at the buyers' side: Since buyers are generally constrained by their limited budgets, signing forward contracts with a large number of sellers can impose over-budget risks, particularly when a large fraction of long-term sellers have attended a transaction. For instance, in Transaction 3 of Fig. 2, buyer $b_3$ should pay all the long-term sellers $s_4-s_7$ for their services, although it may not afford with its budget. 
\end{enumerate}

Thus, existing risks in dynamic and uncertain MCS networks should be well managed to ensure the effectiveness and sustainability. Then, the implementation of spot trading of iFAST confronts the following challenges.

\noindent
$\bullet$~\textit{Selection of volunteers}. A volunteer refers to a buyer who is unable to obtain services offered by long-term sellers. Instead, it can recruit temporary sellers during the transaction. Since overbooking and the limited/uncertain resource supply of long-term sellers can lead to service failures, a proper selection mechanism should be designed to determine appropriate volunteers in each practical transaction when long-term sellers fail to meet the service demand of their contractual buyers. For instance, buyer $b_2$ in Transaction 2 of Fig. 2 is a volunteer of seller $s_4$, since $s_4$'s limited resources and capability can only support two of the contractual buyers (e.g., buyers $b_1$ and $b_3$). 
More importantly, the designed volunteer selection also has to consider diverse requirements of heterogeneous sensing tasks of different buyers (e.g., data quality, transmission time, data size), while ensuring fairness. Common methods for volunteer selection are first-come-first-serve (FCFS, achieving fairness) and greedy-based selection (e.g., buyers with the best channel qualities are volunteers).

\subsection{Step-by-Step Procedure and A Toy Example}
The data trading procedure in our iFAST is detailed below: first, sellers and buyers are encouraged to negotiate proper forward contracts and a feasible overbooking rate by taking into account all the uncertain factors, e.g., resource demand/supply, local workload of sellers, communication conditions among participants. This is implemented before future transactions (see the grey box on the left-side of Fig. 2). Then, in each practical transaction, attendant long-term sellers offer data services to their buyers according to the pre-signed contracts, while those buyers with unsatisfying service qualities are engaged to recruit temporary workers (see the white boxes on the right-side of Fig. 2). 

Fig. 3 depicts a toy example of buyer-seller interactions in iFAST upon considering Transaction 2 in Fig. 2. In particular, sellers $s_1-s_7$ are long-term workers hired prior to the transaction. However, only $s_2$, $s_4$, and $s_5$ can serve the buyers since $s_3$ and $s_7$ are located inside a PoI with no service requests; and $s_1$ and $s_6$ are located outside of all PoIs. In this example, $s_2$, $s_4$, and $s_5$ will offer sensing services to buyers while receiving payments according to their pre-signed forward contracts. However, since seller $s_4$ is faced with heavy local workload and limited resource supply, $b_2$ is selected as a volunteer, which makes $b_2$ fail to meet its expected service quality. To address the possible unsatisfying trading experience, $s_8$ is determined as a temporary worker for the corresponding task of $b_2$ after a certain period of time on spot trading decision-making. This example further highlights the necessity of overbooking in the market. For example, considering buyer $b_1$, the overall resources offered by $s_2$ and $s_4$ can be sufficient to meet its service requirement. However, $b_1$ has signed a contract with 2 extra sellers $s_1$ and $s_3$, in case that $s_1$ and $s_3$ may not be able to join the future transactions. 
 
\section{Case Study}

\noindent
We next show how iFAST can be implemented in practice through a simple mathematical case study. 

\subsection{Outline and Basic Modeling}
We consider a data trading market with one PoI, multiple buyers and sellers, where a buyer can recruit multiple sellers, while each seller can serve at most one task during a transaction to facilitate analysis. A buyer $b_n$ carries a periodic task with a certain desired data quality under a predefined budget. Besides, a seller $s_m$ can contribute its data to buyer $b_n$ with either $q_{m,n}^+$ or $q_{m,n}^-$, as two service quality levels, where the former indicates that $s_m$ gives $b_n$ the highest priority by promising a fixed value of quality, while the latter allows the seller to handle its own local workload first, leading to a fluctuant services. Moreover, participating in different tasks can impose various costs on sellers due to factors such as task’s complexity and distance, for which we introduce $c_{m,n}$ to represent the inherent cost incurred by processing a sensing task. Apparently, different service quality levels can help sellers get different payments. 
To capture the dynamics, we involve the following random variables: the uncertain attendance of sellers $x_m$ that follows a Bernoulli distribution ($x_m=1$ indicates seller $s_m$ attends the transaction); and the varying local workload $l_m$ of each seller that obeys a truncated gaussian distribution. Note that $l_m$ will definitely impact $q_{m,n}^-$ and the practical service cost. For example, a large $l_m$ will lead to a lower $q_{m,n}^-$ when $s_m$ handles its local tasks first, while a larger actual cost (larger than $c_{m,n}$) when the seller offers $q_{m,n}^+$.




According to the above considerations, the utility of a buyer is subsequently defined as its obtained service quality, while that of a seller is given by the difference between its received payment and cost on data services. 

\subsection{Addressing the Key Challenges of iFAST} 
We next demonstrate how the key challenges of iFAST (discussed in Section II. B) are addressed in this case study. First, the design of forward contracts and the overbooking rate can be formulated as a multi-objective optimization (MOO) problem $\bm{\mathcal{F}}$, aiming to maximize the expected overall utility of both buyers and sellers, since the corresponding expectations covers the uncertainties from a long-term view. Besides, each buyer considers two risks: \textit{i)} the risk of receiving an unsatisfying utility (e.g., absence of long-term sellers), which is defined by the probability of the received service quality being less than the requirement; and \textit{ii)} the over-budget risk, indicating the probability that the total payment made to the sellers who have offered services being larger than the buyer's budget, caused by overbooking. Similarly, each seller faces the risk of obtaining a negative utility, which is captured via the probability that the total monetary income being less than the incurred cost of task execution. 
Accordingly, $\bm{\mathcal{F}}$ contains multiple constraints to make all the aforementioned risks within given ranges.

Apparently, solving the MOO problem is non-trivial due to the complexity of sub-objectives (e.g., we should figure out both discrete and continuous unknowns such as the assignment between sellers and buyers, and the trading prices), and the existence of probabilistic constraints (i.e., risks). Thus, we first transform $\bm{\mathcal{F}}$ into a single objective optimization (SOO) problem, by keeping the maximization of buyers' expected utility as the objective, and turn the other one into a constraint, via $\epsilon$-constrained method~\cite{9,13}. This transformation simplifies the problem by assigning a fixed service price to each seller, supporting a non-negative expected profit. Consequently, this SOO challenge is framed as a 0-1 integer programming (IP) problem, demonstrating NP-hardness. To tackle it, we adopt the idea of geometric programming (GP)~\cite{14}, and propose a successive convex algorithm (SCA), where suboptimal solutions are then obtained via commercial convex optimization tools, e.g., CVX. Although the solution design involves a series of approximations, it works well for a variety of problem sizes. 

Then, a spot trading mode is introduced as a backup plan to map buyers with unsatisfying service quality to temporary sellers during each transaction. Specifically, FCFS method is adopted for volunteer selection to achieve fairness. 

\subsection{Evaluations}

Key parameters rely on \textit{i)} real-world dataset of Chicago taxi trips\cite{15}, where the $77^{\text{th}}$ community region is of our interest (PoI). We select some taxis as sellers ($15,16,20$ sellers), while a certain number of tasks ($5,8,10$ buyers) are randomly distributed.
Also, $x_m$ depends on the attendance of these taxis in the considered region during one month, while $c_{m,n}$ can be derived from the motions of taxis, e.g., the distance between a taxi and a buyer. And \textit{ii)} manual parameter settings related to economic behaviors, e.g., $q_{m,n}^+ \in [4,5]$, $q_{m,n}^-=q_{m,n}^+-\xi l_m$ where $\xi\in[0.3,0.5]$ measures
the marginal performance degradation rate, $l_m\sim N(2.5,0.5)$ with different truncate intervals, $c_{m,n}\in [1,1.5]$, the budget and desired service quality of each buyer falls in $[8,10]$ and $[7.5,8.5]$, risks are controlled within $[30\%,40\%]$.

Moreover, since our iFAST is among the first to show a hybrid data delivery mode for MCS, several benchmark methods are considered under conventional spot trading mode: \textit{i)} Spot data delivery (SpotDataD), which utilizes our designed GP-SCA according to the information of each practical transaction; \textit{ii)} Improved implicit enumeration (ImproveIE), as a state-of-the-art method in solving 0-1 IP problem; \textit{iii)} Quality-preferred method (QualityPrefer) where sellers with larger service quality are prioritized; \textit{iv)} Monte Carlo random mapping (MCRandom), representing a common benchmark in many existing works, with good time efficiency. Without loss of generality, simulations are conducted via testing 300 transactions.




\begin{figure}[h!t]
\centering
\includegraphics[width=0.9\linewidth]{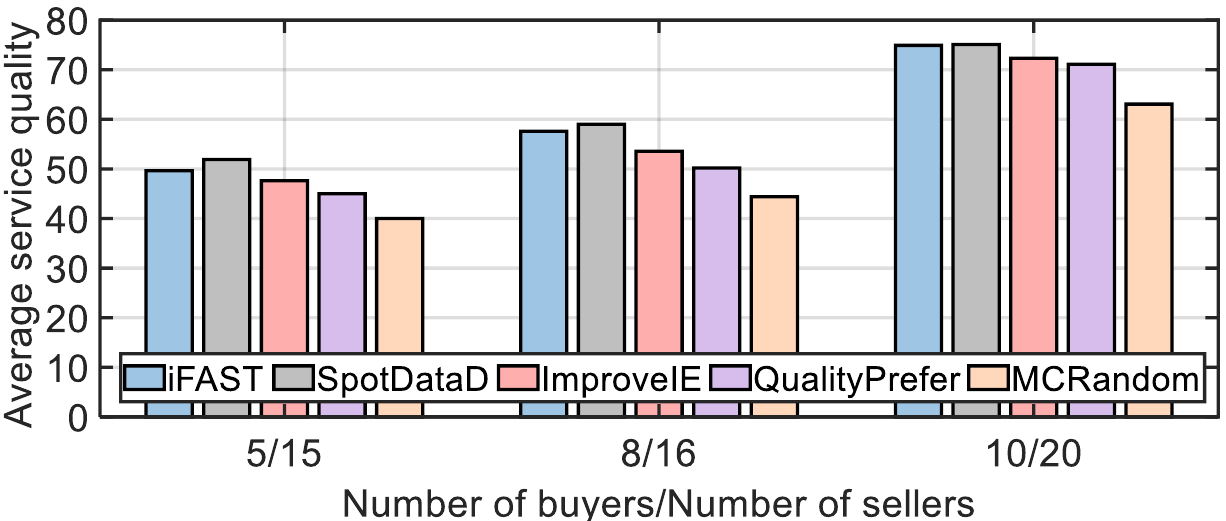}
\caption{Performance evaluation of service quality upon having different numbers of buyers and sellers.}
\end{figure}

\begin{figure}[t]
\centering
\includegraphics[width=0.9\linewidth]{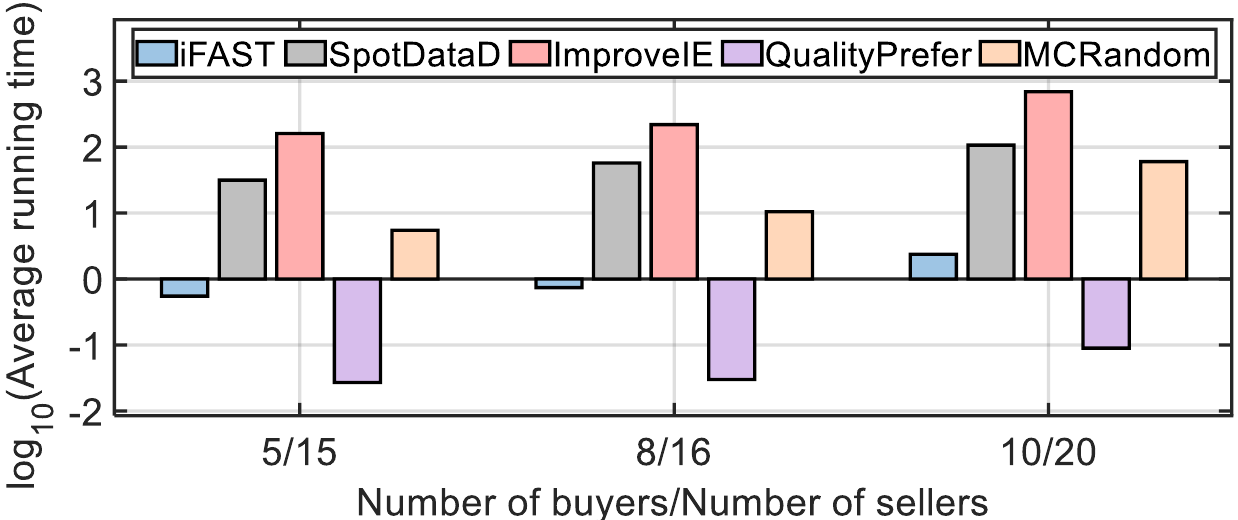}
\caption{Performance evaluation of running time upon having different numbers of buyers and sellers.} 
\end{figure}

We first demonstrate that iFAST obtains comparable performance on service quality comparing to SpotDataD, while outperforming other benchmarks. Then, we show that iFAST achieves such performance under a significantly lower time overhead. 

In Fig. 4, we observe that the average service quality (per transaction) for buyers using our iFAST is marginally lower than that of SpotDataD, since SpotDataD's primary objective is to identify the optimal solution by analyzing the current network/market conditions, whereas iFAST confronts uncertainties and associated risks. Nevertheless, SpotDataD is prone to significant long-term latency due to its reliance on onsite analysis/communications. Moreover, iFAST surpasses other benchmark methods, a testament to the efficacy of our designed overbooking and risk control strategies.
Fig. 5 compares the average running time over 300 transactions, employing log-based representation to highlight the gaps. Notably, iFAST demonstrates a significantly reduced running time, benefiting from the implementation of pre-signed forward contracts, which eliminate the need for buyers and sellers to negotiate trading agreements during each transaction. Conversely, the running times for SpotDataD and ImproveIE exhibit a marked increase with the growth of sellers and buyers. This escalation is attributed to the latency required for each buyer-seller pair to conclude a trading decision, underscoring the efficiency challenges as market participant numbers rise.


In summary, iFAST enables responsive and high-quality data services under significantly lower decision-making overhead as compared to baseline methods, making it a good reference for future large-scale and uncertain MCS networks. 

\section{Inspiration for Future}

\noindent
iFAST introduces a series of open research directions:

\noindent
$\bullet$ \textit{Smart contract design:} To better capture the dynamics of resource trading market, smart forward contracts with flexible terms (e.g., dynamic and environment-aware rights/obligations) rather than fixed ones can be designed via artificial intelligence-based approaches. For example, using deep reinforcement learning via observing the historical statistics of the market. 

\noindent
$\bullet$ \textit{Intelligent risk management:} The forward trading market generally stands for a dynamic coexistence of risks and opportunities, where proper estimation and precise controlling of possible risks facilitate a stable and healthy resource trading market. Thus, it is critical to design a risk management mechanism with intelligent features (e.g., intelligent risk estimation and prediction) rather than the probabilistic approach (where the risks are estimated based on conventional probability theory), to better understand market dynamics.


\noindent
$\bullet$ \textit{Multi-modal resource trading:} Large-scale MCS networks often comprise a set of diverse and heterogeneous resources with different modalities across the sellers (e.g., spectrum, computing, storage, and power). One future direction is to consider the modality of resources in the hybrid resource trading scheme. For example, resources with certain modalities (e.g., storage) can be acquired through forward contracts and are allowed to be overbooked, while the rest of the resources are acquired via spot trading.

\noindent
$\bullet$ \textit{Competition and cooperation among participants:} In a real-world MCS market, competition and cooperation among participants will exist. It is interesting to investigate the possible cooperation and revenue transfer mechanism among sellers (e.g., cooperative game). Also, the cooperation among the buyers can be formed via obtaining the correlation among the tasks executed by them, where the data provided by some sellers acquired by one buyer can be shared across a set of buyers with similar tasks of interest.

\noindent
$\bullet$ \textit{Interference management:} Upon having multiple sellers sending data to the same buyer, effective interference cancellation techniques are needed. More generally, physical layer designs (e.g., power allocation and latency minimization) of the proposed hybrid resource trading market is an open research direction.

\noindent
$\bullet$ \textit{Modeling of diverse uncertainties:}
Modeling the dynamics of arrival and departure of sellers in each PoI (e.g., via probabilistic models) and the rate of service request from each buyer as well as their impact on contract and overbooking rate design is worth further investigation. This direction may lead to an interconnection between reliability analysis of task execution and resource trading in MCS networks.

\noindent
$\bullet$ \textit{Importance-based seller selection:} In addition to their various sensing and computing capabilities, sellers may offer different qualities of data due to their locations. Characterizing the impact of data quality on the execution of crowdsensing tasks and designing proper seller selection mechanism according to data quality is of particular interest.


%
%
%
%
%

\section{Conclusion}
\noindent
We proposed a novel hybrid data service trading mechanism for MCS networks called iFAST, integrating both forward and spot trading modes. We also introduced overbooking and promoted its utilization. Challenges faced with the implementation of iFAST, such as contract term design and risk management, were carefully investigated. A case study was presented along with simulations, revealing that iFAST can achieve commendable performance gains in terms of service quality and time efficiency as compared to conventional mechanisms. Various future research directions were discussed. 

\vfill


\begin{thebibliography}{1}

\bibitem{1} Y. Cui, H. Ding, L. Zhao and J. An, ``Integrated Sensing and Communication: A Network Level Perspective,'' \textit{IEEE Wireless Commun.}, vol. 31, no. 1, pp. 103-109, 2024.

\bibitem{2} F. Wu, S. Yang, Z. Zheng, S. Tang, and G. Chen, ``Fine-Grained User Profiling for Personalized Task Matching in Mobile Crowdsensing,'' \textit{IEEE Trans. Mobile Comput.}, vol. 20, no. 10, pp. 2961-2976, 2021.

\bibitem{3} H. Qi, M. Liwang, S. Hosseinalipour, X. Xia, Z. Cheng, X. Wan,g and Z. Jiao, ``Matching-Based Hybrid Service Trading for Task Assignment Over Dynamic Mobile Crowdsensing Networks,''  \textit{IEEE Trans. Services Comput.}, pp. 1-1, early access, 2023.

\bibitem{4} S. Yang, X. Wang, U. Adeel, C. Zhao, J. Hu, X. Yang, and J. McCann ``The Design of User-Centric Mobile Crowdsensing with Cooperative D2D Communications,'' \textit{IEEE Wireless Commun.}, vol. 29, no. 1, pp. 134-142, 2022.

\bibitem{5} B. Gu, W. Hu, S. Gong, Z. Su and M. Guizani, ``CBDTF: A Distributed and Trustworthy Data Trading Framework for Mobile Crowdsensing,'' \textit{IEEE Trans. Veh. Tech.}, vol. 73, no. 3, pp. 4207-4218,  2024.

\bibitem{6} Z. Zheng, S. Yang, J. Xie, F. Wu, X. Gao, and G. Chen, ``On Designing Strategy-Proof Budget Feasible Online Mechanisms for Mobile Crowdsensing With Time-Discounting Values,'' \textit{IEEE Trans. Mobile Comput.}, vol. 21, no. 6, pp. 2088-2102, 2022.

\bibitem{7} B. Simon, K. Keller, A. Sterz, B. Freisleben, O. Hinz, and A. Klein, ``A Multi-Stakeholder Modeling Framework for the Techno-Economic Analysis of Telecommunication Networks,'' \textit{IEEE Commun. Mag.}, vol. 61, no. 2, pp. 52--56, 2023.

\bibitem{8} S. Sheng, R. Chen, P. Chen, X. Wang, and L. Wu, ``Futures-based Resource Trading and Fair Pricing in Real-Time IoT Networks,'' \textit{IEEE Wireless Commun. Lett}., vol. 9, no. 1, pp. 125-128, 2020.

\bibitem{9} M. Liwang, Z.Gao, and X. Wang, ``Let’s Trade in the Future! A Futures-Enabled Fast Resource Trading Mechanism in Edge Computing-Assisted UAV Networks,'' \textit{IEEE J. Selected Areas in Commun.}, pp. 3252-3270, 2021.

\bibitem{10} M. Dai, Z. Luo, Y. Wu, L. Qian, B. Lin and Z. Su, ``Incentive Oriented Two-Tier Task Offloading Scheme in Marine Edge Computing Networks: A Hybrid Stackelberg-Auction Game Approach,''~\textit{IEEE Trans. Wireless Commun.}, vol. 22, no. 12, pp. 8603-8619, 2023.

\bibitem{11} G. Gao, H. Huang, M. Xiao, J. Wu, Y. e. Sun, and Y. Du, ``Budgeted Unknown Worker Recruitment for Heterogeneous Crowdsensing Using CMAB,'' \textit{IEEE Trans. Mobile Comput.}, vol. 21, no. 11, pp. 3895-3911, 2022.

\bibitem{12} C. Sexton, N. Marchetti and L. A. DaSilva, ``On Provisioning Slices and Overbooking Resources in Service Tailored Networks of the Future,'' \textit{IEEE/ACM Trans. Netw.}, vol. 28, no. 5, pp. 2106-2119, 2020.

\bibitem{13} Zhang, A. K. Qin, W. Shen, L. Gao, K. C. Tan, and X. Li, ``$\epsilon$-Constrained Differential Evolution Using an Adaptive $\epsilon$-Level Control Method,'' \textit{IEEE Trans. Syst., Man, Cybern., Syst.}, pp. 1-17, 2020.

\bibitem{14} M. Ogura, M. Kishida, and J. Lam, ``Geometric Programming for Optimal Positive Linear Systems,'' \textit{IEEE Trans. Auto. Control}, vol. 65, no. 11, pp. 4648-4663, 2020.

%
\bibitem{15} “Taxi trips of chicago in 2013” [Online]. Available: https://data.cityofchicago.org/Transportation/Taxi-Trips2013/6h2x-drp2

\end{thebibliography}
\end{document}